\documentclass[a4paper,11pt]{article} 

\usepackage{amsmath,amsfonts,amssymb} 
\usepackage{graphics,graphicx}				  

\usepackage[colorlinks=false]{hyperref}
\usepackage{latexsym}
\usepackage{cancel}    		 	   
\usepackage{slashed}
\usepackage[T1]{fontenc}
\usepackage{authblk}
\usepackage{mathtools}                

\title{Symmetry Breaking and Proton Decay in Spectral Pati-Salam Model}
\author[1]{Hosein Karimi \thanks{hmk35@mail.aub.edu}}
\affil[1]{Physics Department, American University of Beirut, Lebanon}
\date{}

\begin{document}
	
	\maketitle

\begin{abstract}
	We show the scalar potential in the spectral Pati-Salam model does not provide a suitable vacuum to break to the standard model. We see this potential is proton decay free although diquark and leptoquark vertices exist.  
\end{abstract}
	
	\tableofcontents

\section{Introduction}
	Noncommutative geometry has shown interesting results in physics when it is applied to a hyperspace which is the direct product of four dimensional Riemannian spacetime and a noncommutative space characterized by the spectral triple: algebra of operators which are spacetime functions on the matrices of the form $\mathbb{C} \oplus \mathbb{H} \oplus M_{3}\left(\mathbb{C}\right)$, the relevant Hilbert space of this algebra, and a special operator which generalizes the notion of conventional Dirac operator. The inverse of this operator is looked at as the ultimate propagator between fermions which contains all the useful information of generalized geometry that is the source of the four fundamental forces. Therefore this operator gives dynamic to the fermions in a usual Dirac action way. Hence, all the fluctuations of the Dirac operator in principle connect fermions together. These (classical) fluctuations make the operator invariant under unitary automorphisms of the hyperspace algebra and show themselves as spacetime connections, scalars, and gauge fields. When we try to go beyond the standard model, this property of the Dirac operator puts serious constraints on the scalars in theory. Moreover, these scalars and vector fields along with the spacetime geometry gain dynamic by the spectral action principle which states that their kinetic and potential terms are in the spectrum of the Dirac operator \cite{Chams-1997}. This principle, therefore, dictates the scalar and gauge sectors. For the gauge sector, with the correct choice of algebra, favorable Yang-Mills type action appears \cite{Chams-2010, Chams-2013}. For the scalar sector, the constraints coming from these two different parts of the action are typically conflicting. In the case of spectral standard model these are consistent and reduce the number of free parameters. For a generalized model however we need to make sure of their implications. 
	
	One specific property of the Higgs sector in noncommutative geometry is that Higgses which originate from offdiagonal elements of the self-adjoint Dirac operator of the noncommutative part of hyperspace, connect particles with antiparticles and violate flavor. This is due to the settings of the theory which incorporate antiparticles and particles and treat them quite equally. What is called the first order condition can restrict this phenomenon to happen only for the neutrinos \cite{Chams-2007}. This will then be the source of their Majorana masses. 
	
	Still, there are spaces which do not admit the first order condition like quantum spheres \cite{Dabrowski:2004wr, Dabrowski:2002yb}. Authors of \cite{Chams-20132} have shown that with adding the correct nonlinear terms to the fluctuations of the Dirac operator, $A_{(2)}$ bellow, noncommutative geometry approach can be used for these spaces as well. Invariant physical Dirac operator of the whole space is then (refer to \cite{Chams-2010, Chams-2013, Chams-20132} for exact definitions)
\begin{align}\label{eq-d}
	D_A=D+A_{(1)}+JA_{(1)}J^{-1}+A_{(2)}.
\end{align}
	They also have shown that if Dirac operator can admit first order condition for a subalgebra, starting with it will cause fields in $A_{(2)}$ to be products of fields in $A_{(1)}$ and eat them as their vacuum remnants. The same authors have subsequently used these facts to build a Pati-Salam model based on the algebra $ \mathbb{H} \oplus\mathbb{H} \oplus M_{4}\left(\mathbb{C}\right)$ in \cite{Chams-2013}. Here we consider symmetry breaking scenarios and proton decay possibilities for this model.
	
	Relaxing the first order condition and writing the spectral action based on a broader algebra can cause two main challenges. Firstly, the nonzero offdiagonal sectors of the Dirac operator cause particles other than neutrinos to connect to their antiparticles and produce diquarks which were previously absent in the model. These along with leptoquark scalars on offdiagonal blocks of the diagonal sectors might lead to proton or nucleon decay. Secondly, since the spectral action dictates the scalar potential, it should be checked that the potential can break the symmetry to the standard model in an acceptable way. This is especially important in noncommutative geometry approach because here one does not have the free hand in manipulating the scalar sector which is the case in grand unified or effective field theories. In those cases, if there is a Higgs with correct representation, like a Higgs in the adjoint representation, one may often safely assume that there is a potential for it which provides just enough needed Goldstone bosons. Here however one needs to make sure of that by considering the vacuum of the given potential.

\section{Symmetry Breaking in Spectral Pati-Salam Model}\label{sec-pss}
	In noncommutative geometry approach, unlike usual grand unified models, our hands are not open to choose scalars and customize symmetry breaking along with fermionic mass spectrum. The Higgses and their representations are prescribed by the algebra in the spectral triple and the form of scalar-scalar interactions is dictated by the spectral action principle. This is thought of as one of the advantages of the theory and is also the exact reason that it is highly restrictive and predictive.
	
	In \cite{Chams-2013}, and later on in \cite{Chamseddine:2015ata}, three specific versions of Pati-Salam model are derived and discussed using noncommutative geometry approach. First, the authors start with a Dirac operator which satisfies the first order condition for the standard model subalgebra of $ \mathcal{A}=\mathbb{H} \oplus\mathbb{H} \oplus M_{4}\left(\mathbb{C}\right)$. Then the fluctuations violate this condition. Here, the Higgs representations split due to the way the Dirac operator is written in the beginning. This is called the composite model. Further, they introduce a more general case, called fundamental model, in which the Higgs fields sit in the general representations introduced by the Dirac operator blocks. This Dirac operator is only required to satisfy the axioms (ref. to \cite{Chams-200709}) and not the first order condition at any stage. The Higgs sector of the fundamental model may or may not contain left-right symmetry. Starting with the general case
	
	\begin{align}\label{eq-genericdirac}
	D_A=\left(
	\begin{matrix}
	\left(
	\begin{matrix}
	\slashed{\nabla}_L & \gamma^5\Sigma^{\dot{a}I}_{bJ}\\
	\gamma^5\bar{\Sigma}^{aI}_{\dot{b}J}&\slashed{\nabla}_R
	\end{matrix}
	\right)
	& 
	\left(
	\begin{matrix}
	\gamma^5{H}_{dLcK}&0\\
	0&\gamma^5\dot{H}_{\dot{d}L\dot{c}K}
	\end{matrix}
	\right)\\
	\left(
	\begin{matrix}
	\gamma^5\bar{H}^{dLcK}&0\\
	0&\gamma^5\bar{\dot{H}}^{\dot{d}L\dot{c}K}
	\end{matrix}
	\right) 
	&
	\left(
	\begin{matrix}
	\bar{\slashed{\nabla}}_L&\gamma^5{\Sigma}^{\dot{a}I}_{bJ}\\
	\gamma^5{\bar{\Sigma}}^{aI}_{\dot{b}J}&\bar{\slashed{\nabla}}_R
	\end{matrix}
	\right)
	\end{matrix}\right),
	\end{align}
	the model suggests the following Yukawa interactions and scalar potential in Minkowski spacetime:
	\begin{align} \label{eq-lag}
	L_Y&=g\bar{\psi_R}\Sigma\psi_L + g\bar{\psi_R^C}\dot{H}\psi_R + g\bar{\psi_L^C}H\psi_L + h.c. , \\	\nonumber
	V(\dot{H},\Sigma)&=-\frac{1}{2}M^2 Tr\left( \dot{H}^2 + \Sigma^2 + H^2 \right) \\ \nonumber
	&+ \frac{1}{4}g^2 Tr\Big( 2 \dot{H}^4+ 2 H^4 + 4 |\bar{\dot{H}} \Sigma|^2 + 4 |\bar{H} \Sigma|^2 + |\bar{\Sigma}\Sigma|^2 \\ \nonumber 
	&+ 4\left( \bar{H} \Sigma \dot{H} \bar{\Sigma} + h.c.\right) \Big).
	\end{align}
	
	To study possible symmetry breaking scenarios, $H$ does not help since it cannot get VEV at high energies. Therefore our considerations work for the fundamental models with or without left-right symmetry in the scalar sector. Moreover, we will see in the following that this potential cannot break in a proper way at high energies. Since the fundamental model is supposed to contain the composite model as a special case, this shows that the composite model is not reliable for implying not legitimate breaking scenarios. As an example, there is no proper breaking scenario to split a Higgs in $(15,2,2)$ of $(4C,,2R,2L)$ into a $(15,1,1)$ and a $(1,1,1)$. In this view, when the noncommutative geometry approach is chosen to go beyond the standard model, one needs to make sure the most general form of the scalar sector is chosen and no preconditions are imposed at the algebraic level on the Dirac operator components except the mathematical axioms which are necessary to make the definitions well defined and consistent. Any other preconditions imply spontaneous symmetry breaking scenarios which must be backed by the potential and not be imposed by hand. 
	
	Here the only possible scenario is to break $SU(4)$ and $SU_R(2)$ groups at once using te neutral element of $\dot{H}$. After the breaking, the $36$ degrees of freedom in $\dot{H}$ appear as the following scalars. 
		
\begin{align}\label{eq-particle}
	(10,3,1)_{422}&=6^{-\frac{2}{3}}+6^{\frac{1}{3}}+6^{\frac{4}{3}}+3^{\frac{2}{3}}+3^{-\frac{1}{3}}+3^{-\frac{4}{3}}+ 1^{0}+1^{-}+1^{--}, \\ \nonumber
	(6,1,1)_{422}&=3_A^{\frac{1}{3}}+\bar{3}_A^{-\frac{1}{3}}.
\end{align}
	On the right hand sides, $3$ stands for color triplet, when $\bar{3}$ and $6$ are for antisymmetric and symmetric $3\times3$ representations respectively. Superscripts are the electric charges. The $3$ fields are leptoquarks leading to $\Delta L=\Delta B=1$ processes, while the $\bar{3}_A$ and $6$ fields are diquarks leading to $\Delta B=2$ (figure \ref{fig-vert}). In \cite{Aydemir-2018}, the role of $\bar{3}_A$ as the diquark candidate responsible for the observed B-decay anomalies is discussed and it is noted that this field does not couple with diquarks and does not lead to proton decay processes. Unification of the gauge couplings and the intermediate scale are also addressed in \cite{Chamseddine:2015ata} and \cite{Aydemir:2015nfa}. 
	
	Fields in $\Sigma$ can be written with respect to the standard model quantum numbers as well:
\begin{align}\label{eq-sigma}
	(15,2,2)_{422}&=
	\left(
	\begin{matrix}
	8^0\\
	8^{-}
	\end{matrix}
	\right)_{\frac{1}{2}}
	+
	\left(
	\begin{matrix}
	8^{+}\\
	8^{0}
	\end{matrix}
	\right)_{-\frac{1}{2}}
	+
	\left(
	\begin{matrix}
	\chi_1^0\\
	\chi_1^{-}
	\end{matrix}
	\right)_{\frac{1}{2}}
	+
	\left(
	\begin{matrix}
	\chi_2^{+}\\
	\chi_2^{0}
	\end{matrix}
	\right)_{-\frac{1}{2}}		
	+
	\left(
	\begin{matrix}
	3^{\frac{1}{3}}\\
	3^{-\frac{2}{3}}
	\end{matrix}
	\right)_{\frac{1}{2}}
	+
	\left(
	\begin{matrix}
	3^{-\frac{2}{3}}\\
	3^{-\frac{5}{3}}
	\end{matrix}
	\right)_{-\frac{1}{2}}  \\ \nonumber
	&+
	\left(
	\begin{matrix}
	\bar{3}^{\frac{5}{3}}\\
	\bar{3}^{-\frac{2}{3}}
	\end{matrix}
	\right)_{\frac{1}{2}}
	+
	\left(
	\begin{matrix}
	\bar{3}^{\frac{2}{3}}\\
	\bar{3}^{-\frac{1}{3}}
	\end{matrix}
	\right)_{-\frac{1}{2}}, \\ \nonumber
	(1,2,2)_{422}&=		
	\left(
	\begin{matrix}
	\phi_1^0\\
	\phi_1^{-}
	\end{matrix}
	\right)_{\frac{1}{2}}
	+
	\left(
	\begin{matrix}
	\phi_2^{+}\\
	\phi_2^{0}
	\end{matrix}
	\right)_{-\frac{1}{2}}	.			
\end{align}
	These are left doublets with the noted hypercharges. Fine tuning is required since uncolored elements in trace of $\Sigma$, presented by $\phi$, will have to survive to low energies to make a double Higgs effective action while the colored ones have to gain high masses. We see in the next section that the Higgs sector of the model is proton decay free at the tree level which suggests the tuning might not be severe, yet these scalars couple with fermions and that can put serious restrictions on the intermediate scale. The field $\chi$ is just like standard model Higgs, but only interacts with the leptons.
		
	To have a proper breaking to the standard model, the potential in (\ref{eq-lag}) and its vacuum need to satisfy the following expectations at the tree level. One of the components of the complex field $1^0$ needs to get a VEV and break $(4C,2R,2L)$ directly to $(3C,2L,1Y)$. At the same time, it will be providing Majorana mass for the sterile neutrino due to its Yukawa interaction. The other component of $1^0$ along with two components of $1^-$ have correct quantum numbers to be Goldstone bosons needed for breaking $SU_R(2)$. The six degrees of freedom in the color triplet $3^{\frac{2}{3}}$ and its conjugate should also come massless to play the role of Goldstone bosons for making the six unwanted Higgs fields in $SU(4)$ massive. Moreover, the potential has to provide positive masses proportional to the unification scale for all the other components of $\dot{H}$ and $\Sigma$ since they all are colored or charged and also have Yukawa interactions with the fermions. With the above Higgs content, this scenario is the only possibility to break symmetry to the standard model group which must be implemented automatically by the scalar potential. Specifically, here one cannot break the right and color symmetries at different scales.
	
	Since $H$ does not acquire a VEV, to study the symmetry breaking of the potential in (\ref{eq-lag}), we concentrate only on $\dot{H}$ and $\Sigma$. After normalizing the kinetic terms, in terms of particles in (\ref{eq-particle}), quadratic and quartic parts are
\begin{align}\label{eq-lag2}
		-\frac{1}{2}M^2 \Big(&|6^{-\frac{2}{3}}|^2+|6^{\frac{1}{3}}|^2+|6^{\frac{4}{3}}|^2+|3^{-\frac{4}{3}}|^2+|3^{-\frac{1}{3}}|^2+|3^{\frac{2}{3}}|^2+|1^0|^2+|1^{-}|^2+|1^{--}|^2 \\ \nonumber
		&+|3_A|^2+|\bar{3}_A|^2 + \Sigma^2 \Big) \\ \nonumber
		+\frac{1}{4}g^2 
		\Big(&2|1^0|^4+2|1^{--}|^4+2|3_A|^4+|1^-|^4+|3^{-\frac{4}{3}}|^4+|3^{\frac{2}{3}}|^4+\frac{1}{2}|3^{-\frac{1}{3}}|^4+\frac{1}{2}|\bar{3}_A|^4 \\ \nonumber
		+&4|1^0|^2(|1^-|^2+|3^{\frac{2}{3}}|^2+|3_A|^2+\frac{1}{2}|3^{-\frac{1}{3}}|^2+\frac{\sqrt{2}}{2}(3_A 3^{\frac{1}{3}} + h.c.) \Big) + ...
\end{align}
	Handy calculation shows that this potential can provide needed massless Goldstones. However components of $3_A$ will also appear as six unwanted massless bosons and the rest of scalars will acquire negative masses which means this is a local maximum. To explore a little further, one can easily see that the more general case of
\begin{align}
-\frac{1}{2}M^2 Tr(\bar{\dot{H}} \dot{H}) + \frac{1}{4}\lambda_1 Tr(|\bar{\dot{H} }\dot{H}|^2) + \frac{1}{4}\lambda_2 Tr(|\bar{\dot{H}} \dot{H}|)^2,
\end{align}
	does not work either. The leptoquarks $3^{\frac{2}{3}}$ in the symmetric part with $I_R=1$ and $3_A$ in the antisymmetric part with $I_R=0$ have always the same masses. This is because the underlying group theory causes them to interact with $|1^0|^2$ in the exact same way. Therefore symmetry of the vacuum is bigger than what is needed and any useful vacuum leads to pseudo-Goldstone bosons. If there was a freedom to choose the coefficients, by either $0<-2\lambda_1<\lambda_2$ or $0<2\lambda_1<-\lambda_2$ one could get rid of negative masses and arrange a suitable local minimum (refer to \cite{Li:1973mq} for more). 
	
	The other scalar, $\Sigma$, evidently cannot improve the situation since it is not allowed to have expectation values at high scales. The spectral action mixes $\Sigma$ with $\dot{H}$ which is favorable to make the unwanted scalars massive, but there is a need for fine tuning which is hard to arrange especially if the first breaking cannot happen safely in reasonably low energies. In the literature, the Higgs sector of Pati-Salam is usually thought of as the remnants of $126$ and $120$ or $45$ of $SO(10)$. The last two ones have a $(15,1,1)$ in them and the $126$ has a $(10,3,1)$. So breaking of the right symmetry and the four-color can happen independently \cite{DiLuzio:2011my, Rajpoot:1980xy, Aulakh:1982sw, Babu:2012iv, Mohapatra:1986uf, Langacker:1980js}. 
	
	Hence, lack of terms such as $(Tr(\dot{H}^2))^2$ and $Tr(\dot{H}^2)Tr(\Sigma^2)$ is preventing the potential to have a suitable vacuum. More importantly, current settings of the noncommutative geometry approach cannot offer Higgses in the adjoint representation \cite{Krajewski:1996} and that is problematic for symmetry breaking procedure as we saw here and was seen in \cite{Sars-2011}. As was noted before, it is possible to eliminate Dirac operator elements on offdiagonal blocks in order to have adjoint Higgses. However these would be remnants of forbidden symmetry breakings. Even if the breaking itself was legal, it should be backed by spectral action and not be done in the algebraic level. Obviously, once the most general form of Pati-Salam model is obtained, the only reliable scenario to go down to the standard model is a proper symmetry breaking scenario. Otherwise, the renormalizability of the model is in jeopardy and survived Higgses are not trustable. 
	
\subsection{fermion masses}
	Yukawa terms in \ref{eq-lag} show that $\dot{H}$ only transforms right handed particles to right handed antiparticles while $\Sigma$ transforms right (left) handed particles to left (right) handed particles. 
	
	Each one of $\phi$ and $\chi$ in (\ref{eq-sigma}) provides different Dirac masses for fermions with isospin $\frac{1}{2}$ and $-\frac{1}{2}$ by admitting two independent VEVs. However $\chi$ only couples with the leptons while $\chi$ couples with all the fermions. Neutrino has seesaw mechanism of type one. If $H$ also exists, active neutrino gets a Majorana mass as well and seesaw mechanism of type two happens naturally for the neutrino. 
	
	With four independent VEVs and Majorana masses for the neutrinos, there is enough freedom and the model basically does not have predictions for fermion masses. It is also interesting that in the offdiagonal part of generic Dirac operator, naturally a colorless double Higgs arises which gives masses to all the particles. A restriction however emerges when the quartic part of potential is not able to provide high masses for the components of $8$ in $\Sigma$, which is just like what happens for the $6$ in $\dot{H}$. Through $\Sigma^4$, only $\chi$ can give a mass term to the $8$ which will inevitably be from the same order of quarks masses. Therefore $\Sigma$ will come with an overall negative mass term. The problem is originating from tight restrictions of two parts of the action. Bosonic part which does not possess all the possible quartic terms due to the spectral action principle, and the fermionic part which couples all the scalars to fermions.
	
	This model basically predicts Yukawa and gauge couplings to be equal to the unified gauge coupling at unification. They of course will run differently. Even if we implement a procedure to produce Yukawas as $3$ by $3$ matrices and find some room for maneuver, the scalar couplings will not be independent of Yukawas and the $g$. These two originate from constants in the Dirac operator of the discrete space and the cutoff function in the spectral action which are the only sources to produce couplings in noncommutative geometry approach. The scalar couplings therefore are expected to be from the same order of these couplings at unification. It is seen that the relation between these couplings are consistent with particle masses at low energies \cite{Chams-2012, Khozani:2017ykq}. Therefore from any aspect it is inevitable to try to find a way for entering new fields into the picture. 
	
	\section{Proton Decay}
	Another interesting test for the potential suggested by the spectral action principle for the Pati-Salam model is to look for proton decay diagrams. The importance of this task in general is to see whether we can safely bring the intermediate breaking mass scale down. This could add to this model the privilege of needing a less severe fine tuning and also can have phenomenological implications for the achievable energies at LHC \cite{Aydemir:2015nfa}. Another encouraging fact is that noncommutative geometry approach tries to yield an effective theory at the unification scale and proton decay is one of the few probes available to examine a theory which is written at those high energies with our current experimental abilities.
	
	In this section we first do a general analysis of the possible proton decay diagrams with the Yukawa interactions of (\ref{eq-yukawa}). Then we look for the needed vertexes of Higgs interactions in the potential. It is enough to look for vertices originating from $\dot{H}^2$ and $\dot{H}^4$ because $\Sigma$ does not lead to diquarks or leptoquarks and $H$ has the same exact Yukawa terms for left handed fermions as $\dot{H}$ has for right handed ones. This feature originates from grading of the algebra and left-right symmetry because each one of the left and right handed particles has a Higgs field to produce flavor violating diagrams. In contrast, what happens in the $SU(5)$ unified theory is that the same extra gauge fields couple with different chiralities. 
	
	If the proton decay diagrams are allowed by potential at tree level, then one should make sure all the elements which are involved get indeed a mass at high enough scales. Then we should also be worried since $H$ cannot admit VEV and doesn't couple with $\dot{H}$ in a way to gain a high mass from it. However it becomes clear in the following that the needed order six and order nine operators cannot form by vertexes that the potential provides. 
	
\begin{figure} 
		\includegraphics[width=0.9\linewidth]{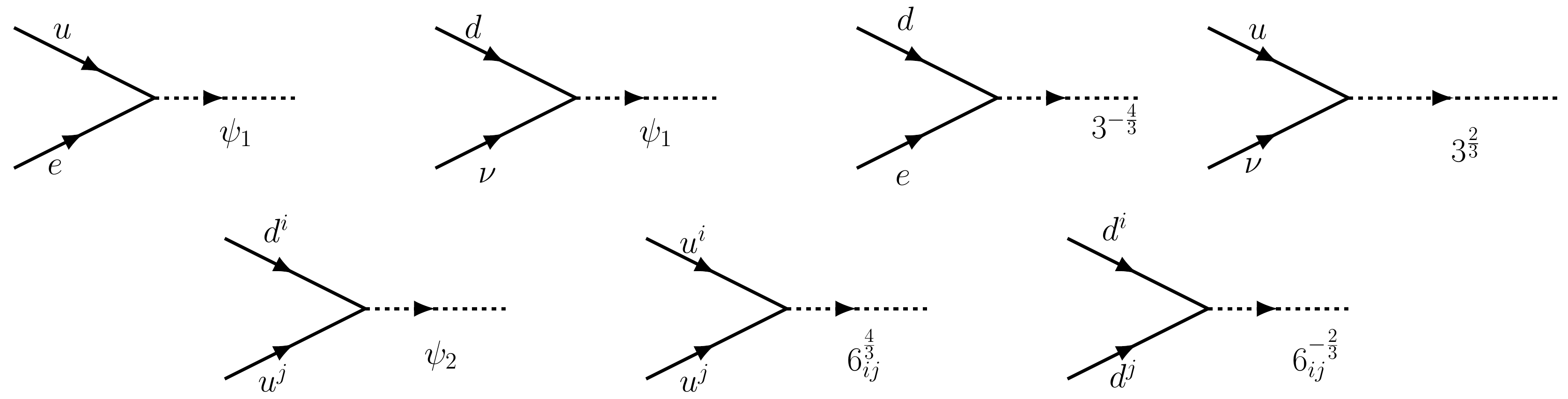}
		\caption{Diquark and leptoquark diagrams. $\psi_1$ is a combination of $3^{-\frac{1}{3}}$ and $3^{-\frac{1}{3}}_A$} and $\psi_2$ is a combination of $\bar{3}_A$ and $6^{\frac{1}{3}}.$
		\label{fig-vert}
\end{figure}
	
	It is evident that the extra gauge fields in this model cannot cause proton decay on their own because diquark vertices do not exist. This roots in the fact that in noncommutative approach all particles sit in the ordinary, rather than ordinary and conjugate, fundamental representations (similar to SO(10) and usual Pati-Salam unified models). Yet the Yukawa terms in (\ref{eq-yukawa}) suggest that diquark scalars exist. We saw in \ref{sec-pss} that $\dot{H}$ alone could not break the symmetry. If that was the case, three fields in $3^{-\frac{2}{3}}$ and their conjugates would have been eaten by gauge bosons and disappeared from Lagrangian. Now however one part of each remains as a Higgs field. We are interested to see whether the Higgs fields can cause proton decay at tree level or not. With the particle content in (\ref{eq-particle}) and their dangerous diagrams in figure \ref{fig-vert} we witness that only following vertexes can cause proton decay diagrams with up to dimension nine operators.
\begin{align}
	&\psi_1 \psi_2, \quad 3^{-\frac{4}{3}} 6^{\frac{4}{3}}, \quad 3^{\frac{2}{3}} 6^{-\frac{2}{3}}, \\ \nonumber
	&\psi_1 \psi_1 3^{\frac{2}{3}}, \quad \psi_2 \psi_2 6^{-\frac{2}{3}}, \quad 3^{\frac{2}{3}} 3^{\frac{2}{3}} 3^{-\frac{4}{3}}, \quad 6^{\frac{4}{3}} 6^{-\frac{2}{3}} 6^{-\frac{2}{3}}.
\end{align}
	Here $\psi_1$ is a combination of $3_A$ and $3^{-\frac{1}{3}}$. Also $\psi_2$ is a combination of $6^{\frac{1}{3}}$ and $\bar{3}_A$. 
	
	Obviously all the terms in $\dot{H}^4$ are from order four, so we seek for the above terms coupled with one or two $1^0$. This field admits a VEV and dimension of diagrams remain intact. However such terms are absent in the potential. All the terms in $\dot{H}$ are presented in appendix \ref{sec-hdot}. For the relevant dimension six operators we read from (\ref{eq-hdot})
	\[|1^0|^2 \left(|3^{\frac{2}{3}}|^2+|3_A|^2+\frac{1}{2}|3^{-\frac{1}{3}}|^2+\frac{\sqrt{2}}{2}(3_A 3^{\frac{1}{3}} + h.c.) \right), \]
	which can only lead to pion decay. The followings are all of the possible dimension nine vertexes. None of these is in the above form and they do not lead to proton decay.
\begin{align}
	1^0 \Big( &3^{-\frac{2}{3}}6^{\frac{4}{3}}3^{-\frac{2}{3}}+ 3^{-\frac{2}{3}}6^{\frac{1}{3}}3^{\frac{1}{3}} + 3^{\frac{1}{3}}6^{\frac{1}{3}}3^{-\frac{2}{3}} + 3^{\frac{1}{3}}6^{-\frac{2}{3}}3^{\frac{1}{3}} \\ \nonumber
	&\qquad\qquad\qquad+ {3_A}^* 6^{-\frac{2}{3}} {3_A}^* + 2 (3^{-\frac{2}{3}} \bar{3}_A {3_A}^*) +3^{-\frac{2}{3}} \bar{3}_A {3_A}^* + h.c.\Big)
\end{align}

\section{Conclusion}
	When the spectral action is used to go beyond the standard model, clearly, the challenge is that the spectral potential has to be able to support a suitable symmetry breaking scenario. With the current settings of noncommutative geometry approach it is not possible to arrange an acceptable scenario as it is seen here and in reference \cite{Sars-2011}. 
	
	The model specially lacks the presence of Higgses in the adjoint representation and Higgses which decouple with the fermions in the action. Both of these are automatically true for Higgses on diagonal blocks of diagonal sectors of Dirac operator. However the algebra is even graded and odd parts of the Dirac operator cannot coexist with its even parts. The only way might be to add a real singlet on diagonal of Dirac operator. This couples with all the Higgses and might be able to provide a suitable minimum and solve the problem of negative mass terms in section \ref{sec-pss}. Anyways, it is essential to expand the geometry to generate more suitable fields and provide enough complexity to have proper symmetry breaking. It is an important fact that spectral potential is in a way that no any Higgs can break the symmetry alone even if it is in the adjoint of the four-color and the right symmetries. Therefore, for example, all the Higgses on diagonal blocks would be needed to break the symmetry in a complicated form.
	
	We also conclude that the above project cannot be done by superficially splitting the Higgses to the needed representations with imposing algebraic conditions on the original discrete Dirac operator. When we work with the most general form of the Dirac operator which is consistent with the mathematical axioms and go beyond the standard model, this is the role of the spectral action principle now to be able to provide a proper potential to break the unified symmetry to the standard model group. It is certainly not legitimate to assume such process is automatically available and manipulate the scalar fields by imposing the first order condition or making any other request on the original, and not perturbed, Dirac operator. At the very least, this puts renormalizability in jeopardy.
	
 	We also considered another important aspect of the spectral potential which is its implications for proton decay. The result is that the offdiagonal Higgses which violate baryonic and leptonic numbers in their Yukawa interactions do not lead to proton decay at tree level up to nine dimensional operators. This is despite the presence of dangerous leptoquark and diquark vertexes.

\section*{Appendix}
\appendix

\section{Terms in $\dot{H}$ with respect to standard model representations}\label{sec-hdot}

	Here we present all terms in $\dot{H}^4$ which is the only part of the potential with diquarks and leptoquarks. Similar terms exist in the $H^4$. Terms are normalized so the kinetic terms be in canonical form.
	
\begin{align}\label{eq-hdot}
	\dot{H}^4&=|1^0|^4+|1^{--}|^4+\frac{1}{2}|1^-|^4+ 2|1^-|^2 \left(|1^0|^2+|1^{--}|^2 \right) \\ \nonumber
	&+ (1^+)^2 1^0 1^{--} + (1^-)^2 {1^0}^* 1^{++} \\ \nonumber
	&+2|1^0|^2 \left(|3^{\frac{2}{3}}|^2+|3_A|^2+\frac{1}{2}|3^{-\frac{1}{3}}|^2+\frac{\sqrt{2}}{2}(3_A 3^{\frac{1}{3}} + h.c.) \right) \\ \nonumber
	&|1^-|^2 \left(|3^{\frac{2}{3}}|^2+2|3_A|^2+|3^{-\frac{1}{3}}|^2+|3^{-\frac{4}{3}}|^2 \right) \\ \nonumber
	&+2|1^{--}|^2 \left(|3_A|^2+\frac{1}{2}|3^{-\frac{1}{3}}|^2+|3^{-\frac{4}{3}}|^2-\frac{\sqrt{2}}{2}(3_A 3^{\frac{1}{3}} + h.c.) \right) \\ \nonumber
	&+(1^+ 1^0 + 1^{++}1^-) \left( 3^{-\frac{2}{3}}3^{-\frac{1}{3}}+ 3^{\frac{1}{3}}3^{-\frac{4}{3}} +\sqrt{2}({3_A}^* 3^{-\frac{4}{3}}+ 3^{-\frac{2}{3}} 3_A) \right) 
\end{align}
\begin{align*}
	&+\frac{1}{2}|3^{-\frac{4}{3}}|^4+ \frac{1}{2}|3^{\frac{2}{3}}|^4+ \frac{1}{4}|3^{-\frac{1}{3}}|^4+ |3^{-\frac{1}{3}}|^4 + |3_A|^4 \\ \nonumber
	&+ \left( |3^{\frac{2}{3}}|^2 + |3^{-\frac{4}{3}}|^2 \right)+ 2\left((3^{\frac{1}{3}})^2 3^{\frac{2}{3}} 3^{-\frac{4}{3}} + h.c.\right)\\ \nonumber
	&+ {3_A}^* 3_A \left( |3^{\frac{2}{3}}|^2 +|3^{-\frac{1}{3}}|^2+ |3^{-\frac{4}{3}}|^2 \right) \\ \nonumber
	&+ \frac{1}{2} \left( 3^{-\frac{2}{3}} 3_A {3_A}^* 3^{\frac{2}{3}} + 3^{\frac{4}{3}} 3_A {3_A}^* 3^{-\frac{4}{3}} + 3^{\frac{1}{3}} 3_A {3_A}^* 3^{-\frac{1}{3}} \right) \\ \nonumber
	&+ \left( 3^{-\frac{2}{3}} \bar{3_A} {\bar{3_A}}^* 3^{\frac{2}{3}} + 3^{\frac{4}{3}} \bar{3_A} {\bar{3_A}}^* 3^{-\frac{4}{3}} + 3^{\frac{1}{3}} \bar{3_A} {\bar{3_A}}^* 3^{-\frac{1}{3}} + {3_A}^* \bar{3_A} \bar{3_A}^* 3_A \right) \\ \nonumber
	&+\sqrt{2} \left( 3^{-\frac{2}{3}} \bar{3}_A {3_A}^* 1^0 + 3^{\frac{4}{3}} \bar{3}_A {3_A}^* 1^{--} + 3^{\frac{1}{3}} \bar{3}_A {3_A}^* 1^- + h.c.\right)
\end{align*}
\begin{align*}
	&+4 Tr \Big( 3^* .6.6^* .3 + {3_A}^*6.6^*3_A + \big( (6^* 3_A). (\epsilon 3^* \epsilon \bar{3_A})  \\ \nonumber
	& \qquad\qquad\qquad\qquad+(3^*.6) \bar{3_A}^* 3_A + \bar{3_A}^* \epsilon.3.3^*.6 - 3^*.6.6^*.\epsilon 3_A +h.c. \big) \Big) \\ \nonumber
	&+|\bar{3}_A|^4 + Tr \Big( |6^*.6|^2+ 2 \bar{3_A}^* 6. 6^* \bar{3}_A + 2 (6^*.6) ({\bar{3}_A}^* \bar{3}_A) \\ \nonumber
	 & \qquad\qquad\qquad\qquad+ \big( 6^*\bar{3}_A.\epsilon 6^* \epsilon \bar{3}_A - 2 6^*.6.6^*.\epsilon +h.c. \big) \Big)\\ \nonumber
	 &+ \sqrt{2} \left( 3^{-\frac{2}{3}} \bar{3}_A {3_A}^* 1^0 + 3^{\frac{1}{3}}\bar{3}_A {3_A}^* 1^- + 3^{\frac{4}{3}}\bar{3}_A {3_A}^* 1^{--}+h.c.\right) \\ \nonumber
	 &+  \Big( {3_A}^* 6^{-\frac{2}{3}} {3_A}^* 1^0+  {3_A}^* 6^{\frac{1}{3}} {3_A}^* 1^- + {3_A}^* 6^{\frac{4}{3}} {3_A}^* 1^{--} +h.c.\Big) \\ \nonumber
	 &+  \Big( 3^{-\frac{2}{3}}6^{\frac{4}{3}}3^{-\frac{2}{3}}+ 3^{-\frac{2}{3}}6^{\frac{1}{3}}3^{\frac{1}{3}} + 3^{\frac{1}{3}}6^{\frac{1}{3}}3^{-\frac{2}{3}} + 3^{\frac{1}{3}}6^{-\frac{2}{3}}3^{\frac{1}{3}} + h.c.\Big)  1^0
\end{align*}

\section*{Acknowledgment}	
	The author is grateful to Professor Ali Chamseddine for useful discussions. He also thanks Professor Khalil Bitar for his generous time spending and insightful discussions.
	
	\bibliography{ref-patisalamprotondecay}{}
	\bibliographystyle{plain}
	
\end{document}